\documentclass[multphys,vecphys]{svmult}

\usepackage{makeidx}         % allows index generation
\usepackage{graphicx}        % standard LaTeX graphics tool
\usepackage{multicol}        % used for the two-column index
\usepackage[bottom]{footmisc}% places footnotes at page bottom
\makeindex             % used for the subject index
\begin{document}

\title*{Heating Rate Profiles in Galaxy Clusters}
% Use \titlerunning{Heating rates} for an abbreviated version of
\author{Edward C. D. Pope\inst{1,2,3}, Georgi Pavlovski\inst{2}, Christian R. Kaiser\inst{2} \and
Hans Fangohr\inst{3}}
\institute{School of Physics and Astronomy, University of Leeds, Leeds, LS2 9JT  
\texttt{e.c.d.pope@leeds.ac.uk} 
\and
School of Physics and Astronomy, University of Southampton
 Highfield, Southampton, SO17 1BJ
\and
School of Engineering Sciences, University of Southampton Highfield,
Southampton, SO17 1BJ}
%\institute{Name and Address of your Institute
%\texttt{name@email.address}
%\and Name and Address of your Institute \texttt{name@email.address}}

\maketitle

\section{Abstract}

The results of hydrodynamic simulations of the Virgo and Perseus
clusters suggest that thermal conduction is not responsible for the
observed temperature and density profiles. As a result it seems that
thermal conduction occurs at a much lower level than the Spitzer
value. Comparing cavity enthalpies to the radiative losses within the
cooling radius for seven clusters suggests that some clusters are
probably heated by sporadic, but extremely powerful, AGN outflows
interspersed between more frequent but lower power outflows.

\section{Introduction}

The two candidates for heating cluster atmospheres are Active Galactic
Nuclei (AGNs) and thermal conduction. Heating by AGN is thought to
occur through the dissipation of the internal energy of plasma bubbles
inflated by the AGN at the centre of the cooling flow. Since these
bubbles are less dense than the ambient gas, they are buoyant and rise
through the intracluster medium (ICM) stirring and exciting sound
waves in the surounding gas. This energy may be dissipated by means of
a turbulent cascade, viscous processes, or aerodynamic forces. Deep in
the central galaxy other processes such as supernovae and stellar
winds will also have some impact on the ambient gas.

Thermal conduction may also play a significant role in transferring
energy towards central regions of galaxy clusters given the
temperature gradients which are observed in many clusters.

\section{The Model}

\subsection{General heating rates}

Starting from the assumption that the atmospheres of galaxy clusters
are spherically symmetric, and in a quasi steady-state, it is
possible, using the fluid energy equation, to derive what the radial
time-averaged heating rate must be in order to maintain the observed
temperature and density profiles. The flow of the gas is assumed to be
subsonic meaning that the cluster atmosphere is in approximate
hydrostatic equilibrium allowing the gravitational acceleration to be
calculated from observations of the temperature and density
profiles. To avoid anomalies when calculating spatial derivatives,
continuous analytical functions are fitted through the density and
temperature data. This ensures that there are not any large
discontinuities which may result in extreme, and erroneous, heating
rates later on in the calculations. This model is described in greater
detail in \cite{pope06}

\subsection{Thermal conduction}

Thermal conduction of energy from the cluster outskirts may provide
the required heating of the central regions without an additional
energy source, like an AGN. The thermal conductivity is assumed to be
given by \cite{spitzer}, but includes a suppression factor designed to
take into account the possible effects of magnetic fields. For a
steady-state to exist, the heating by thermal conduction must equal
the heating rate. From this criterion, the radial suppression factor
can be deduced.

\subsection{Heating by AGNs}

The time-averaged mechanical power of an AGN can be estimated by
dividing the cavity enthalpy by a characteristic timescale, see for
example \cite{birzan}. We assume that the radio-emitting plasma that
fills the cavities is relativistic and that half of the outburst
energy is deposted in the ICM by shocks. An accurate estimate of the
time-averaged jet power requires the average time between consecutive
AGN outbursts to be known. However, since this parameter is rarely the
known, a typical choice would be the buoyant timescale required for
the cavity to rise to its current location. An alternative method is
to assume a particular value for the period of the AGN. In this study
it is initially assumed that the period of each AGN is $10^{8}$ yrs.

Note that this is simply an estimate of the rate at which energy is
injected by the AGN and is not related to any particular physical
process by which this energy is dissipated, e.g. the viscous
dissipation of sound waves.

An estimate of the period required to balance the radiative losses
within the cooling radius is obtained by calculating the volume
integral of the heating rate within this region and comparing this
with the bubble enthalpy.

\section{Results:1}

Radial suppression factors and AGN periods are calculated for a sample
of seven objects for which temperature and density were available, as
well as the information about the X-ray cavities inflated by their
central AGNs. These objects are the Virgo \index{virgo}
\cite{ghizzardi04}, Perseus \index{perseus} \cite{sanders04} and Hydra
\index{hydra} \cite{davhyd01} clusters, A2597 \index{A2597}
\cite{mcnam01}, A2199 \index{A2199}, A1795 \index{A1795} \cite{ettori}
and A478 \index{A478} \cite{sun} with cavity parameters taken from
\cite{birzan}.

\subsection{Suppression factors}

The results show that the suppression factors must be finely tuned if
thermal conduction is to balance the radiative losses. Such a high
degree of fine-tuning suggests that thermal conduction is unlikely to
be a dominant heating mechanism in galaxy clusters. Furthermore, in
many cases, the required suppression factors exceed the physical
maxmimum of unity. This is most true for the Virgo \index{virgo},
Hydra \index{hydra} and A2597 \index{A2597} clusters. In contrast, it
appears that thermal conduction could, in principle, balance the
radiative losses in the Perseus \index{perseus}, A2199 \index{A2199},
A478 \index{A478} and probably A1795 \index{A1795}. The effect of
thermal conduction on a cluster from each of these two groups is
investigated in more detail using numerical simulations discussed in
the next section.

\subsection{AGN Duty Cycles}

The periods for Virgo \index{Virgo} and A478 \index{A478} are of the
order of $10^{6}$ yrs which is very short compared to the predicted
lifetimes of AGN \cite{nipbin}. In contrast, the Hydra cluster
requires recurrent outbursts of magnitude similar to the currently
observed one only every $10^{8}$ yrs, or so. The required duty cycles
for the remaining AGNs are of the order of $10^{7}$ yrs. From this,
the obvious conclusion is that if thermal conduction is negligible and
if this sample is representative of galaxy clusters in general, then
many, if not all, clusters will probably be heated, at certain points
in time, by extremely powerful AGN outbursts. Furthermore, it is
worthwhile pointing out that roughly 71\% of cD galaxies at the
centres of clusters are radio-loud \cite{burns} which is larger than
for galaxies not at the centres of clusters. This may suggest that the
galaxies at the centres of clusters are indeed active more frequently
than other galaxies.

\section{The Simulations}\label{sims}

Numerical simulations of mock Virgo and Perseus clusters were
performed using the FLASH hydrodynamics code. Four simulations were
performed for each cluster to investigate the effect of different
values of the thermal conduction on the evolution of the cluster
temperature and density profiles. The simulations of the Virgo cluster
are described in more detail in \cite{pope05}.

\section{Results:2}

\subsection{Temperature and Electron Number Density Profiles} \label{res1}

\begin{figure*} 
\begin{minipage}[b]{.5\linewidth}
\centering\includegraphics[width=0.8\linewidth]{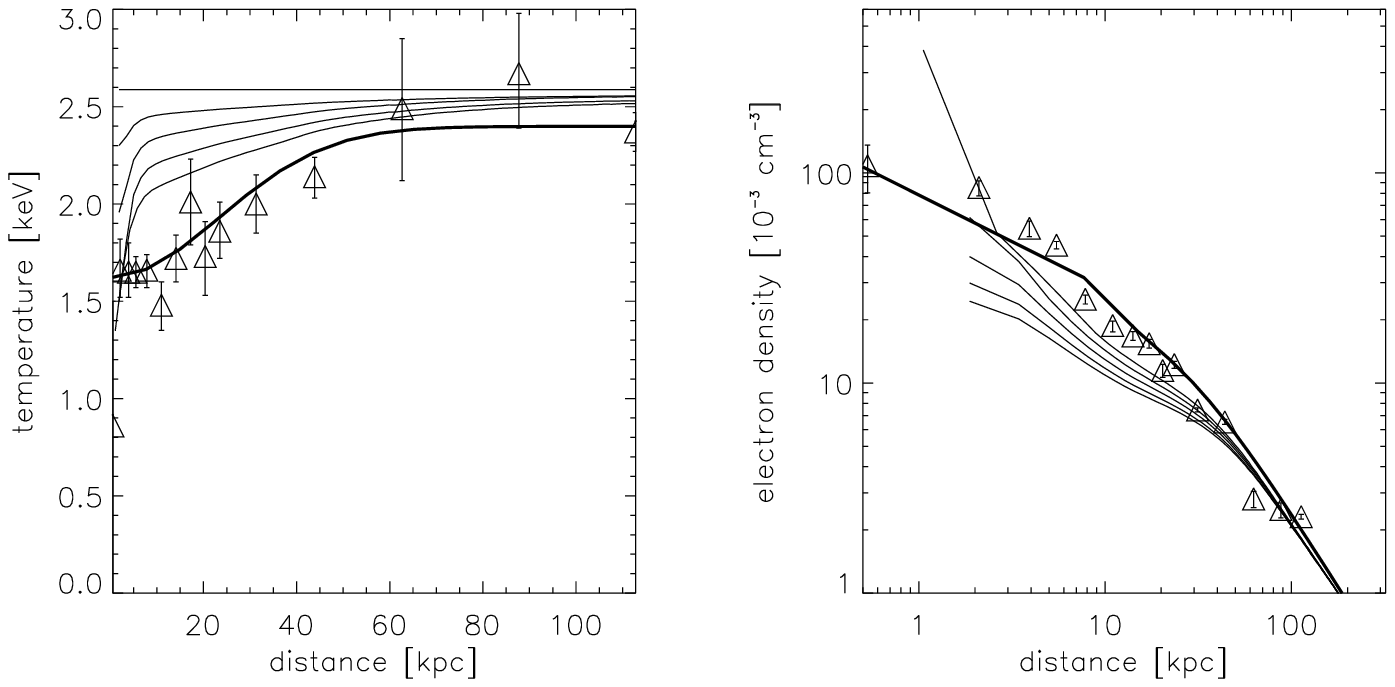}
\end{minipage}\hfill

\begin{minipage}[b]{.5\linewidth}
\centering\includegraphics[width=0.8\linewidth]{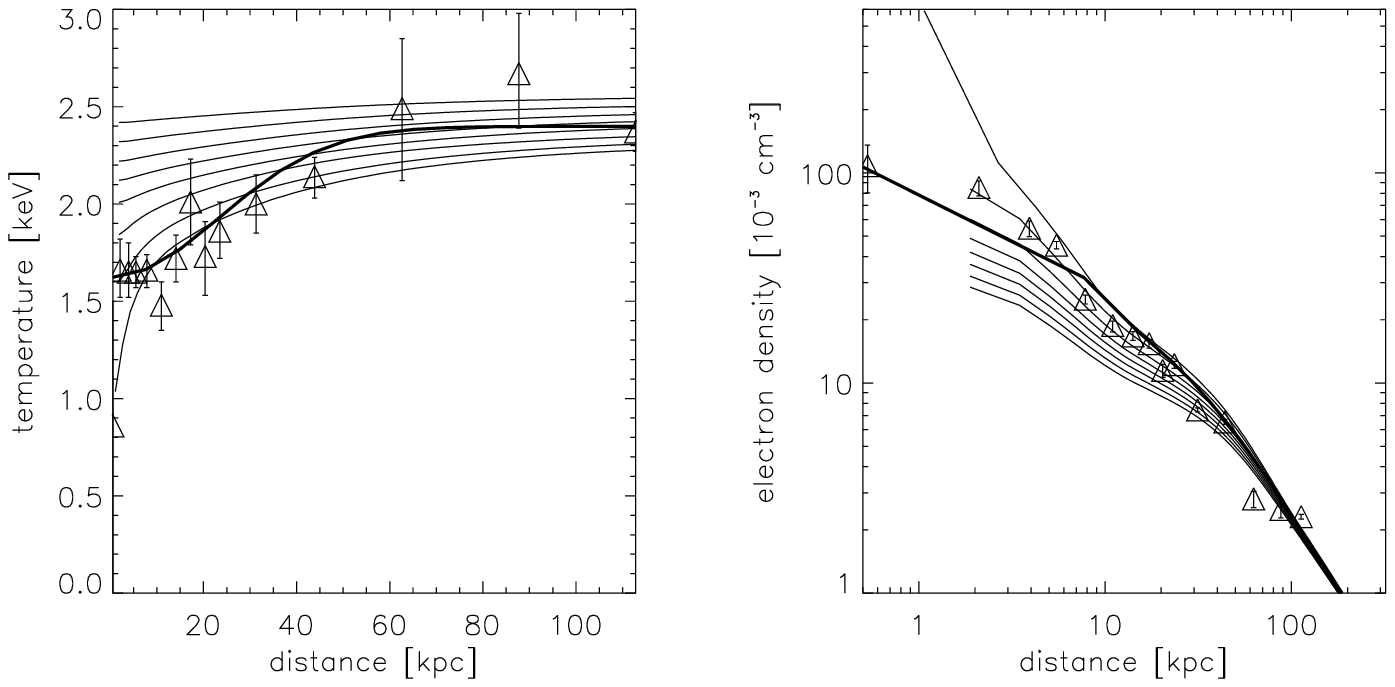}
\end{minipage}\hfill
\caption{Temperature and density profiles evolving with time for zero
thermal conduction (top) and Spitzer thermal conduction (bottom). The
thick lines for both temperature and density are the functions fitted
to the data points (triangles) by \cite{ghizzardi04}.  For zero
thermal conduction the top line in the temperature plot shows the
temperature profile after $3.17 \times 10^{8} $ yr and the bottom line
at time of the end of the simulation. The intermediate lines represent
the temperatures at intervals of $ 3.17 \times 10^{8} $ yr after the
top temperature profile. The temporal sequence of the lines is
reversed (bottom to top) in the density plot. For Spitzer thermal
conduction the data are plotted every $6.34 \times 10^{8} $
yrs.}\label{fig:0sp}
\end{figure*}

For the Virgo cluster, the temperature and density data are presented
as spherically averaged profiles, rather than 1-d slices through the
cluster. The temperatures and densities for two cases are compared to
the observations of \cite{ghizzardi04} in figure \ref{fig:0sp}. The
two cases shown here are: zero thermal conduction and Spitzer thermal
conduction.

\begin{figure*} 
\begin{minipage}[b]{.5\linewidth}
\centering \includegraphics[width=0.8\linewidth]{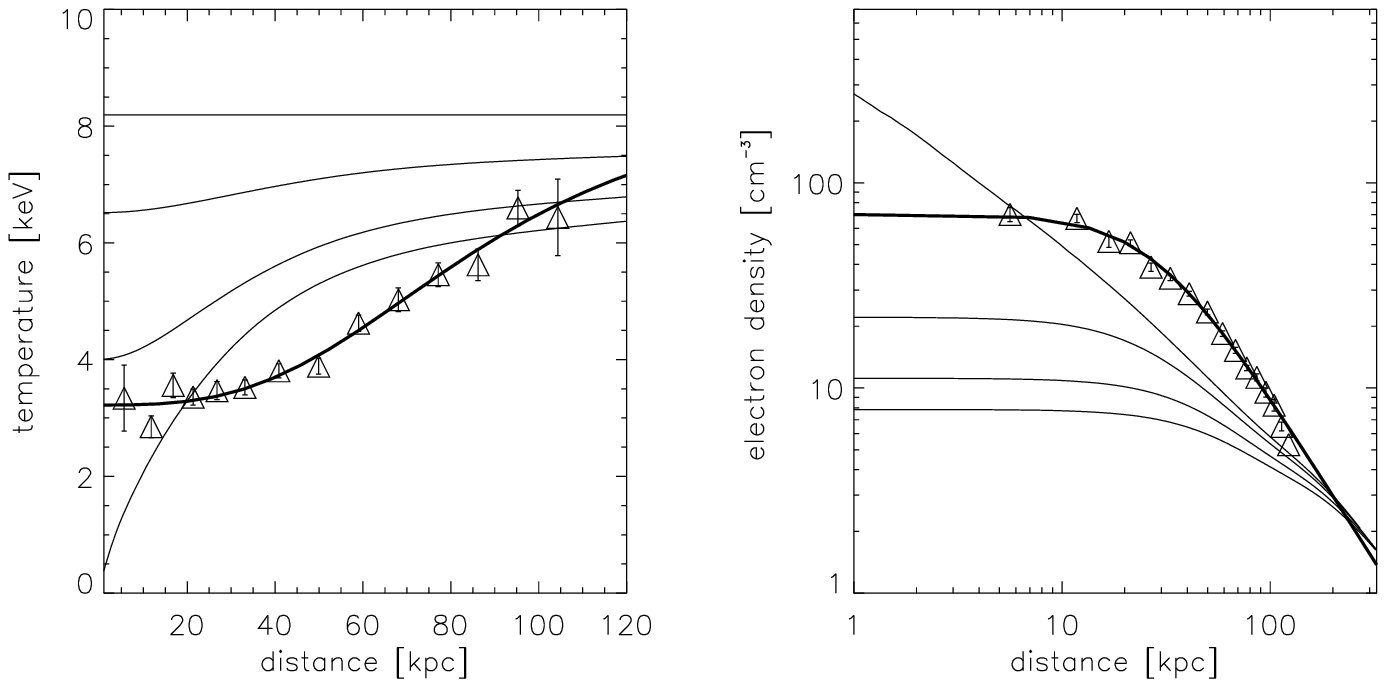}
\end{minipage}\hfill

\begin{minipage}[b]{.5\linewidth}
\centering \includegraphics[width=0.8\linewidth]{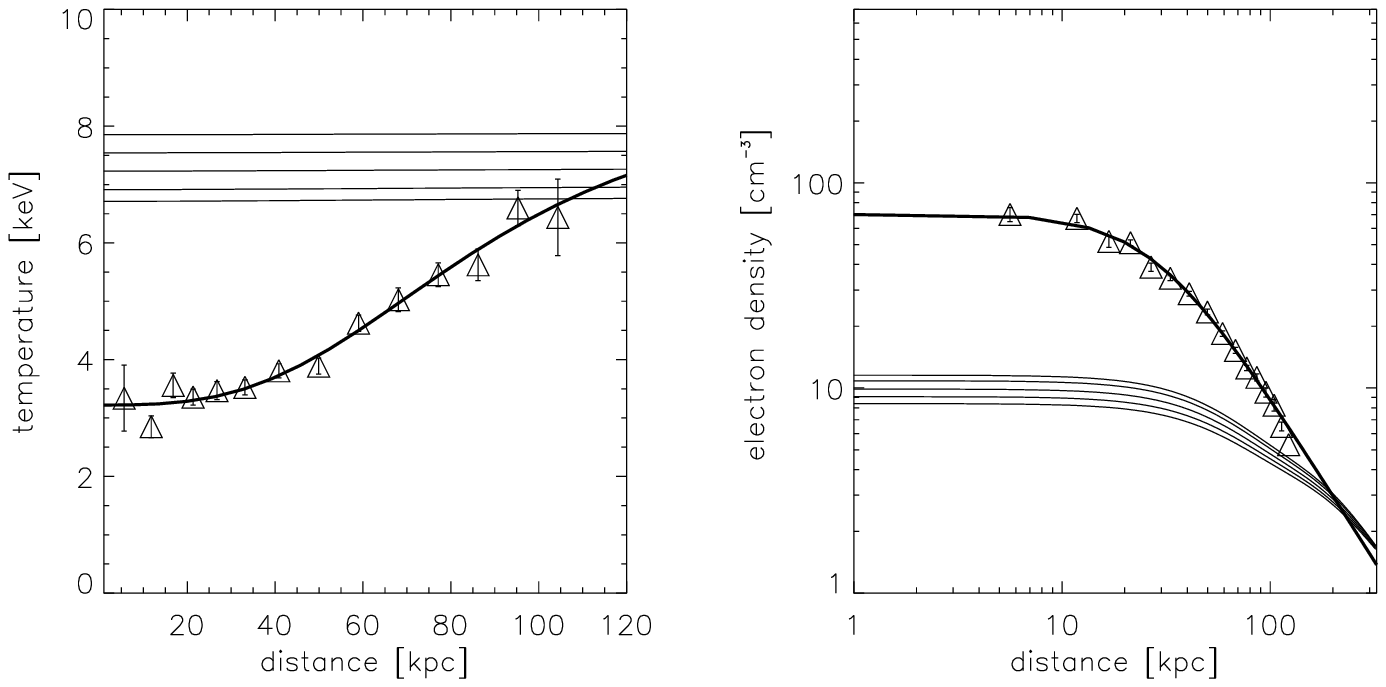}
\end{minipage}\hfill
\caption{Temperature and density profiles evolving with time for zero
thermal conduction (top) and Spitzer thermal conduction (bottom). The
thick lines for both temperature and density are the functions fitted
to the data points (triangles) by \cite{sanders04}.  The top line in
the temperature plot shows the temperature profile after $6 \times
10^{9} $ yr and the bottom line at time of the end of the
simulation. The intermediate lines represent the temperatures at
intervals of $ 3 \times 10^{9} $ yrs after the top temperature
profile. The temporal sequence of the lines is reversed (bottom to
top) in the density plot.}\label{fig:p0sp}
\end{figure*}

The simulations of the Perseus cluster were 1-d, but spherically
symmetric, meaning that the data did not require spherical
averaging. Results are shown in figure \ref{fig:p0sp} for the same
cases as the Virgo cluster.

The qualitative response of the two clusters to the presence of
thermal conduction is rather similar: it seems that thermal conduction
probably cannot indefinitely prevent a cooling catastrophe from
occuring. This is characterised by a large dip and a peak in the
temperature and density profiles, respectively. 

In the case of the Virgo cluster, even full Spitzer thermal conduction
can only postpone the cooling catastrophe for a few Gyrs. The result
is roughly the same for simulations of the Perseus cluster where the
thermal conduction is sub-Spitzer. The main difference is that in the
Perseus cluster the time taken for a cooling catastrophe to develope
is significantly longer than for the Virgo cluster. To some extent
this is because the density of the gas is lower in Perseus but also
because the energy transport by thermal conduction is greater, due to
the higher gas temperatures.

In the Perseus cluster, the simulations show that including full
Spitzer thermal conduction can avert a cooling catastrophe, for at
least a Hubble time, by transfering energy at such a high rate that it
essentially keeps the temperature profile flat. This also prevents the
density profile from evolving significantly. However, the rapid
transfer of energy by thermal conduction leads to an additional
problem: thermal conduction `heats' the central regions of galaxy
clusters by transfering thermal energy from outskirts yet there is not
an infinite supply. Thus, while energy is being transfered the average
gas temperature drops. Eventually the entire ICM would cool to
temperatures where it is no longer observed in the X-rays. Essentially
the problem with thermal conduction is that it does not `add' energy
to a system, it merely transfers it from one region to another.

\section{Conclusion}

The results of simulations including thermal conduction qualitatively
agree with predictions, based on energetic arguments, for determining
which clusters will be most influenced by thermal conduction. However,
for clusters such as Perseus, where thermal conduction can balance the
radiative losses, the simulated temperature profile never converges
with the observations. The results of these simulations suggest that
thermal conduction must be drastically reduced compared to the Spitzer
value.

Current techniques of estimating AGN power output are still relatively
uncertain. Nevertheless, the values obtained using these methods
sometimes require periods as short as a few Myrs to balance the
radiative losses. An alternative possibility is that the AGN power
output is currently underestimated, meaning that the period is
actually longer. Another strong possibility is that extemely powerful
outbursts occur between more frequent but less powerful outbursts.

%
%
% BibTeX users please use
\bibliographystyle{plain}
\bibliography{database}

\begin{thebibliography}{10}

\bibitem{birzan}
L.~{B{\^ i}rzan}, D.~A. {Rafferty}, B.~R. {McNamara}, M.~W. {Wise}, and
  P.~E.~J. {Nulsen}.
\newblock {A Systematic Study of Radio-induced X-Ray Cavities in Clusters,
  Groups, and Galaxies}.
\newblock {\em ApJ}, 607:800--809, June 2004.

\bibitem{burns}
J.~O. {Burns}.
\newblock {Numerical Models of Extragalactic Radio Sources: An Observer's
  Perspective}.
\newblock {\em Bull. Am. Astron. Soc.}, 22:821--+, March 1990.

\bibitem{davhyd01}
L.~P. {David}, P.~E.~J. {Nulsen}, B.~R. {McNamara}, W.~{Forman}, C.~{Jones},
  T.~{Ponman}, B.~{Robertson}, and M.~{Wise}.
\newblock {A High-Resolution Study of the Hydra A Cluster with Chandra:
  Comparison of the Core Mass Distribution with Theoretical Predictions and
  Evidence for Feedback in the Cooling Flow}.
\newblock {\em ApJ}, 557:546--559, August 2001.

\bibitem{ettori}
S.~{Ettori}, A.~C. {Fabian}, S.~W. {Allen}, and R.~M. {Johnstone}.
\newblock {Deep inside the core of Abell 1795: the Chandra view}.
\newblock {\em MNRAS}, 331:635--648, April 2002.

\bibitem{ghizzardi04}
S.~{Ghizzardi}, S.~{Molendi}, F.~{Pizzolato}, and S.~{De Grandi}.
\newblock {Radiative Cooling and Heating and Thermal Conduction in M87}.
\newblock {\em ApJ}, 609:638--651, July 2004.

\bibitem{mcnam01}
B.~R. {McNamara}, M.~W. {Wise}, P.~E.~J. {Nulsen}, L.~P. {David}, C.~L.
  {Carilli}, C.~L. {Sarazin}, C.~P. {O'Dea}, J.~{Houck}, M.~{Donahue},
  S.~{Baum}, M.~{Voit}, R.~W. {O'Connell}, and A.~{Koekemoer}.
\newblock {Discovery of Ghost Cavities in the X-Ray Atmosphere of Abell 2597}.
\newblock {\em ApJ}, 562:L149--L152, December 2001.

\bibitem{nipbin}
C.~{Nipoti} and J.~{Binney}.
\newblock {Time variability of active galactic nuclei and heating of cooling
  flows}.
\newblock {\em MNRAS}, 361:428--436, August 2005.

\bibitem{pope05}
E.~C.~D. {Pope}, G.~{Pavlovski}, C.~R. {Kaiser}, and H.~{Fangohr}.
\newblock {The effects of thermal conduction on the intracluster medium of the
  Virgo cluster}.
\newblock {\em MNRAS}, 364:13--28, November 2005.

\bibitem{pope06}
E.~C.~D. {Pope}, G.~{Pavlovski}, C.~R. {Kaiser}, and H.~{Fangohr}.
\newblock {Heating rate profiles in galaxy clusters}.
\newblock {\em MNRAS}, 367:1121--1131, April 2006.

\bibitem{sanders04}
J.~S. {Sanders}, A.~C. {Fabian}, S.~W. {Allen}, and R.~W. {Schmidt}.
\newblock {Mapping small-scale temperature and abundance structures in the core
  of the Perseus cluster}.
\newblock {\em MNRAS}, 349:952--972, April 2004.

\bibitem{spitzer}
L.~{Spitzer}.
\newblock {\em Physics of Fully Ionized Gases}.
\newblock Wiley-Interscience, New York, 1962.

\bibitem{sun}
M.~{Sun}, C.~{Jones}, S.~S. {Murray}, S.~W. {Allen}, A.~C. {Fabian}, and A.~C.
  {Edge}.
\newblock {Chandra Observations of the Galaxy Cluster A478: The Interaction of
  Hot Gas and Radio Plasma in the Core, and an Improved Determination of the
  Compton y-Parameter}.
\newblock {\em ApJ}, 587:619--624, April 2003.

\end{thebibliography}
%
% Non-BibTeX users please follow the syntax
% the syntax of "referenc.tex" for your own citations
%\input{referenc}
%%%%%%%%%%%%%%%%%%%%%%%%%%%%%%%%%%%%%%%%%%%%%%%%%%%%%%%%%%%%%%%%%%%%%%  }

%%%%%%%%%%%%%%%%%%%%%%%%%%%%%%%%%%%%%%%%%%%%%%%%%%%%%%%%%%%%%%%%%%%%%%

\printindex
\end{document}